\documentstyle[12pt]{article}
\newcommand\be{\begin{equation}}
\newcommand\ee{\end{equation}}
\newcommand\bea{\begin{eqnarray}}
\newcommand\eea{\end{eqnarray}}
\newcommand\ket[1]{|#1\rangle}

\newcommand\braket[2]{\langle #1|#2\rangle}
\newcommand{\fatalpha}{{\bf \alpha \kern -0.44em \alpha}}
\newcommand{\fatsigma}{{\bf \sigma \kern -0.54em \sigma}}
\newcommand{\tpchi}{{\bf \chi \kern -0.35em \chi}}
\newcommand{\llambda}{{\bf \lambda \kern -0.45em \lambda}}

\renewcommand{\theequation}{\arabic{equation}}
\renewcommand{\theequation}{\thesection-\arabic{equation}}
\bibliography{plain}
\title{\bf Perfect transference of a $d$-level quantum state over pseudo-distance-regular networks}\vspace{20mm}
\author{ M. A. Jafarizadeh$^{a,b,c}$
 \thanks{E-mail:jafarizadeh@tabrizu.ac.ir},
 R. Sufiani$^{a,b}$
 \thanks{E-mail:sofiani@tabrizu.ac.ir}, S. F. Taghavi$^{a}$ and E. Barati$^{a}$
 \\ $^a${\small Department of Theoretical Physics and Astrophysics,
University of Tabriz, Tabriz 51664, Iran.} \\ $^b${\small
Institute for Studies in Theoretical Physics and Mathematics,
Tehran 19395-1795, Iran.} \\ $^c${\small Research Institute for
Fundamental Sciences, Tabriz 51664, Iran.}} \pagebreak

\vspace{20mm}
\begin{document}
\maketitle \vspace{15mm}
\newpage
\begin{abstract}
Following the prescription of Ref. \cite{PST} in which perfect
state transference (PST) of a qubit over distance regular spin
networks was discussed, in this paper PST of an arbitrary
$d$-level quantum state (qudit) over antipodes of more general
networks called pseudo distance-regular networks, is investigated.
In fact, the spectral analysis techniques used in the previous
work \cite{PST}, and algebraic structures of pseudo
distance-regular graphs are employed to give an explicit formula
for suitable coupling constants in the Hamiltonians so that the
state of a particular qudit initially encoded on one site will
evolve freely to the opposite site without any dynamical control,
i.e., we show that how to derive the parameters of the system so
that PST can be
achieved.\\

{\bf Keywords:Perfect state transfer, $d$-level quantum state,
Stratification, Pseudo-distance-regular network}

{\bf PACs Index: 01.55.+b, 02.10.Yn }
\end{abstract}

\vspace{70mm}
\newpage
\section{Introduction}
The transference of quantum information, encoded in a quantum
state, from one part of a physical unit, e.g., a qubit, to another
part is a crucial ingredient for many quantum information
processing protocols \cite{1}. There are various physical systems
that can serve as quantum channels, one of them being a quantum
spin system. Quantum communication over short distances through a
spin chain, in which adjacent qubits are coupled by equal strength
has been studied in detail, and an expression for the fidelity of
quantum state transfer has been obtained \cite{Bose,5}. Similarly,
in Ref. \cite{6}, near perfect state transfer was achieved for
uniform couplings provided a spatially varying magnetic field was
introduced. After the work of Bose \cite{Bose}, in which the
potentialities of the so-called spin chains have been shown,
several strategies were proposed to increase the transmission
fidelity \cite{Os} and even to achieve, under appropriate
conditions, perfect state transfer \cite{8,9'',Bu,Bu1,yung,yung1}.
All of these proposals refer to ideal spin chains in which only
nearest-neighbor couplings are present. In Refs. \cite{8,9''}, the
$d$-dimensional hypercube with $2^d$ vertices has been projected
to a linear chain with $d+1$ sites so that, by considering fixed
but different couplings between the qubits assigned to the sites,
the PST can be achieved over arbitrarily long distances in the
chain. In Ref. \cite{PST}, the so called distance-regular graphs
have been considered as spin networks (in the sense that with each
vertex of a distance-regular graph a qubit or a spin was
associated ) and perfect state transference (PST) over them has
been investigated, where a procedure for finding suitable coupling
constants in some particular spin Hamiltonians has been given so
that perfect transference of a quantum state between antipodes of
the networks can be achieved. The aim of this paper is to extend
this proposal to systems of particles with arbitrary number of
levels, the so-called qudits. These systems can be appeared in
condensed matter and solid state physics such as the fermionic
$SU(N)$ Hubbard model \cite{hubb,hubb1,hubb2}. In Ref.
\cite{karim}, state transference over spin chains of arbitrary
spin has been discussed so that an arbitrary unknown qudit be
transferred through a chain with rather good fidelity by the
natural dynamics of the chain. In this work, we focus on the
situation in which state transference is perfect, i.e., the
fidelity is unity. Furthermore, we consider more general graphs
called pseudo distance regular graphs or $QD$ type graphs
\cite{obata,obh} (distance regular graphs are special kinds of
pseudo distance regular ones) as underlying networks and
investigate PST over antipodes of these networks. In fact this
work is an extension of the previous work \cite{PST} to PST of a
qudit over a pseudo distance regular network. We use the
techniques such as stratification of the graphs
\cite{obata},\cite{js}-\cite{jss1} and spectral distribution
associated with the graphs. Then, we consider particular
hamiltonians with nonlinear terms and give a method for finding a
suitable set of coupling constants so that PST over antipodes of
the networks be possible. In fact, we give an explicit formula for
suitable coupling constants so that PST between the first node of
the networks and the opposite one can be achieved. As examples we
will consider some important pseudo distance regular networks such
as the $G_n$ networks, the modified $G_n$ networks, Hadamard
network, etc.

The organization of the paper is as follows: In section 2, we
review some preliminary facts about graphs and their
stratification, pseudo distance-regular graphs and spectral
distribution associated with graphs. Section $3$ is devoted to PST
of a qudit over antipodes of pseudo distance-regular networks,
where a method for finding suitable coupling constants in
particular spin Hamiltonians so that PST be possible, is given.
The paper is ended with a brief conclusion and one appendix.
\section{Preliminaries}In this section we recall some
preliminaries related to graphs, their stratifications and the
notion of pseudo-distance-regularity (as a generalization of
distance regularity) of graphs.
\subsection{Graphs and their stratifications}
In this section, we review the stratification of the graphs and
the notion of pseudo-distance-regularity.

A graph is a pair $\Gamma=(V,E)$, where $V$ is a non-empty set
called the vertex set and $E$ is a subset of $\{(x,y):x,y \in
V,x\neq y\}$ called the edge set of the graph. Two vertices $x, y
\in V$ are called adjacent if $(x,y)\in E$, and in that case we
write $x \sim y$. For a graph $\Gamma=(V,E)$, the adjacency matrix
$A$ is defined as
\begin{equation}\label{adj.}
\bigl(A)_{\alpha, \beta}\;=\;\cases{1 & if $\;\alpha\sim \beta$
\cr 0 & \mbox{otherwise}\cr}.
\end{equation}
Conversely, for a non-empty set $V$, a graph structure is uniquely
determined by such a matrix indexed by $V$.

The degree or valency of a vertex $x \in V$ is defined by
\begin{equation}\label{val.}
\kappa(x)=|\{y\in V: y\sim x\}|
\end{equation}
where, $|\cdot|$ denotes the cardinality. The graph is called
regular if the degree of all of the vertices be the same. In this
paper, we will assume that graphs under discussion are regular. A
finite sequence $x_0, x_1, . . . , x_n \in V$ is called a walk of
length $n$ (or of $n$ steps) if $x_{i-1}\sim  x_i$ for all $i= 1,
2, . . . , n$. Let $l^2(V)$ denote the Hilbert space of $C$-valued
square-summable functions on $V$. With each $\beta\in V$ we
associate a vector $\ket{\beta}$ such that the $\beta$-th entry of
it is $1$ and all of the other entries of it are zero. Then
$\{\ket{\beta}: \beta\in V\}$ becomes a complete orthonormal basis
of $l^2(V)$. The adjacency matrix is considered as an operator
acting in $l^2(V)$ in such a way that
\begin{equation}
A\ket{\beta}=\sum_{\alpha\sim \beta}\ket{\alpha}.
\end{equation}

Now, we recall the notion of stratification for a given graph
$\Gamma$. To this end, let $\partial(x,y)$ be the length of the
shortest walk connecting $x$ and $y$ for $x\neq y$. By definition
$\partial(x,x)=0$ for all $x\in V$. The graph becomes a metric
space with the distance function $\partial$. Note that
$\partial(x,y)=1$ if and only if $x\sim y$. We fix a vertex $o \in
V$ as an origin of the graph, called the reference vertex. Then,
the graph $\Gamma$ is stratified into a disjoint union of strata
(with respect to the reference vertex $o$) as
\begin{equation}\label{strat}
V=\bigcup_{i=0}^{\infty}\Gamma_{i}(o),\;\ \Gamma_i(o):=\{\alpha\in
V: \partial(\alpha,o)=i\}
\end{equation}
Note that $\Gamma_i(o)=\emptyset$ may occur for some $i \geq 1$.
In that case we have $\Gamma_i(o)= \Gamma_{i+1}(o)=...=
\emptyset$. With each stratum $\Gamma_i(o)$ we associate a unit
vector in $l^2(V)$ defined by
\begin{equation}\label{unitv}
\ket{\phi_{i}}=\frac{1}{\sqrt{\kappa_i}}\sum_{\alpha\in
\Gamma_{i}(o)}\ket{\alpha},
\end{equation}
where, $\kappa_i=|\Gamma_{i}(o)|$ is called the $i$-th valency of
the graph ($\kappa_i:=|\{\gamma:\partial(o,\gamma)=i
\}|=|\Gamma_{i}(o)|$).
\subsection{Pseudo-distance regular graphs}
Given a vertex $\alpha\in V$ of a graph $\Gamma$, consider the
stratification (\ref{strat}) with respect to $\alpha$ such that
$\Gamma_i(\alpha)=\emptyset$ for $i > d$. Then, we say that
$\Gamma$ is pseudo-distance-regular around vertex $\alpha$
whenever for any $\beta \in \Gamma_k(\alpha)$ and $0 \leq k \leq
d$, the numbers
\begin{equation}\label{pseudo}
c_k(\beta):=\frac{1}{\kappa(\beta)}\sum_{\gamma\in
\Gamma_1(\beta)\cap \Gamma_{k-1}(\alpha)}\kappa(\gamma),\;\
a_k(\beta):=\frac{1}{\kappa(\beta)}\sum_{\gamma\in
\Gamma_1(\beta)\cap \Gamma_{k}(\alpha)}\kappa(\gamma),\;\
b_k(\beta):=\frac{1}{\kappa(\beta)}\sum_{\gamma\in
\Gamma_1(\beta)\cap \Gamma_{k+1}(\alpha)}\kappa(\gamma),
\end{equation}
do not depend on the considered vertex $\beta \in
\Gamma_k(\alpha)$, but only on the value of $k$. In such a case,
we denote them by $c_k$, $a_k$ and $b_k$ respectively.

It should be noticed that for regular graphs $\Gamma$
($\kappa(\beta)=\kappa\equiv \kappa_1$ for all $\beta\in V$), the
numbers $c_k,a_k$ and $b_k$ read as
\begin{equation}\label{pseudo'}
c_k=|\Gamma_1(\beta)\cap \Gamma_{k-1}(\alpha)|,\;\
a_k=|\Gamma_1(\beta)\cap \Gamma_{k}(\alpha)|,\;\
b_k=|\Gamma_1(\beta)\cap \Gamma_{k+1}(\alpha)|,
\end{equation}
where we tacitly understand that $\Gamma_{-1}(\alpha)=\emptyset$.
The intersection numbers (\ref{pseudo'}) and the valencies
$\kappa_i=|\Gamma_i(\alpha)|$ satisfy the following obvious
conditions
$$a_i+b_i+c_i=\kappa,\;\;\ \kappa_{i-1}b_{i-1}=\kappa_ic_i ,\;\;\
i=1,...,d,$$
\begin{equation}\label{intersec'}
\kappa_0=c_1=1,\;\;\;\ b_0=\kappa_1=\kappa, \;\;\;\ (c_0=b_d=0).
\end{equation}

One should notice that, the definition of pseudo-distance regular
graphs together with the Eq.(\ref{intersec'}), imply that in
general, the valencies $\kappa_i$ ( the size of the $i$-th
stratum) for $i=0,1,...,d$ do not depend on the considered vertex
$\beta \in \Gamma_k(\alpha)$, but only on the value of $k$.

The notion of pseudo-distance regularity has a close relation with
the concept of QD type graphs introduced by Obata \cite{obata},
such that for the adjacency matrices of this type of graphs, one
can obtain a quantum decomposition associated with the
stratification (\ref{strat}) as
\begin{equation}\label{QD'}
A=A^++A^-+A^0,
\end{equation}
where, the matrices $A^+$, $A^-$ and $A^0$ are defined as follows:
for $\beta\in \Gamma_k(\alpha)$, we set
\begin{equation}\label{QD}
A^+\ket{ \beta}=\sum_{\delta\in \Gamma_{k+1}(\alpha),\\
\delta\sim\beta}\ket{\delta},\;\ A^-\ket{\beta}=\sum_{\delta\in
\Gamma_{k-1}(\alpha),\delta\sim\beta}\ket{\delta},\;\ A^0\ket{
\beta}=\sum_{\delta\in \Gamma_{k}(\alpha),\delta\sim\beta}\ket{
\delta},
\end{equation}
Since $\beta\in \Gamma_k(\alpha)$ and $\beta \sim \delta$ then
$\delta\in \Gamma_{k-1}(\alpha) \cup \Gamma_k(\alpha)\cup
\Gamma_{k+1}(\alpha)$.

It has been shown in Ref. \cite{obata} that, if the closed
subspace of $l^2(V)$ spanned by $\{\ket{\phi_{i}},i=0,\ldots,
d-1\}$ be invariant under the quantum components $A^+,A^-$ and
$A^0$, then there exist two sequences (called Szeg\"{o}- Jacobi
sequences) $\{\omega_l\}_{l=1}^{\infty}$ and
$\{\alpha_l\}_{l=1}^{\infty}$ derived from $A$ such that
$$A^+\ket{\phi_l}=\sqrt{\omega_{l+1}}\ket{\phi_{l+1}},\;\ l\geq0 \;\
,$$
$$ A^-\ket{\phi_0}=0,\;\
A^-\ket{\phi_l}=\sqrt{\omega_l}\ket{\phi_{l-1}},\;\ l\geq 1,$$
\begin{equation}\label{QD1}
A^0\ket{\phi_l}=\alpha_{l}\ket{\phi_l}, l\geq0,
\end{equation}
where $\omega_{l+1} = \frac{\kappa_{l+1}}{\kappa_l} \kappa^2_-(j)$
with $\kappa_-(j) = |\{i\in \Gamma_l(\alpha) :\;\ i \sim j\}|$ for
$j\in \Gamma_{l+1}(\alpha)$, and $\alpha_{l}= \kappa_0(j)=|\{i \in
V_l; i \sim j\}|$ for $j\in \Gamma_l(\alpha)$, for $l\geq0$. One
can easily check that the coefficients $\alpha_l$ and $\omega_l$
are given by
\begin{equation}\label{omegal0}
\alpha_l=\kappa-b_{l}-c_{l},\;\;\;\;\ \omega_l=b_{l-1}c_{l},\;\;\
l=1,...,d.
\end{equation}

By using (\ref{QD'}) and (\ref{QD1}), one can obtain
\begin{equation}\label{QD2}
A\ket{\phi_l}=\sqrt{\omega_{l+1}}\ket{\phi_{l+1}}+\alpha_{l}\ket{\phi_l}+\sqrt{\omega_l}\ket{\phi_{l-1}},\;\
l\geq0.
\end{equation}
One should notice that, the vectors $\ket{\phi_i}, i=0,1,...,d-1$
form an orthonormal basis for the so called Krylov subspace
$K_d(\ket{\phi_0},A)$ defined as
\begin{equation}
K_d(\ket{\phi_0},A) = span\{\ket{\phi_0},A\ket{\phi_0},
\cdots,A^{d-1}\ket{\phi_0}\}.
\end{equation}
Then it can be shown that \cite{krylov}, the orthonormal basis
$\ket{\phi_i}$ are written as
\be\label{phi}\ket{\phi_i}=P_i(A)\ket{\phi_0},\ee where
$P_i=a_0+a_1A+...+a_iA^i$ is a polynomial of degree $i$ in
indeterminate $A$ (for more details see for example
\cite{js,krylov}).

It may be noted that, the pseudo-distance-regularity is a
generalization of the notion of distance-regularity which is
defined as:\\ \textbf{Definition (distance-regular graphs).} A
pseudo-distance regular graph $\Gamma=(V,E)$ is called
distance-regular with diameter $d$ if for all $k\in\{0,1,...,d\}$,
and $\alpha,\beta\in V$ with $\beta\in \Gamma_k(\alpha)$, the
numbers $c_k(\beta)$, $a_k(\beta)$ and $b_k(\beta)$ defined in
(\ref{pseudo}) depend only on $k$ but do not depend on the choice
of $\alpha$ and $\beta$.

Now, it should be noticed that, the stratification of
distance-regular graphs will be independent of the choice of the
reference vertex (the vertex which stratification is done with
respect to it).
\subsection{Spectral distribution of the graphs}
It is well known that, for any pair $(A,\ket{\phi_0})$ of a matrix
$A$ and a vector $\ket{\phi_0}$, one can assign a measure $\mu$ as
follows
\begin{equation}\label{sp1}
\mu(x)=\braket{ \phi_0}{E(x)|\phi_0},
\end{equation}
 where
$E(x)=\sum_i|u_i\rangle\langle u_i|$ is the operator of projection
onto the eigenspace of $A$ corresponding to eigenvalue $x$, i.e.,
\begin{equation}
A=\int x E(x)dx.
\end{equation}
Then, for any polynomial $P(A)$ we have
\begin{equation}\label{sp2}
P(A)=\int P(x)E(x)dx,
\end{equation}
where for discrete spectrum the above integrals are replaced by
summation. Therefore, using the relations (\ref{sp1}) and
(\ref{sp2}), the expectation value of powers of adjacency matrix
$A$ over reference vector $\ket{\phi_0}$ can be written as
\begin{equation}\label{v2}
\braket{\phi_{0}}{A^m|\phi_0}=\int_{R}x^m\mu(dx), \;\;\;\;\
m=0,1,2,....
\end{equation}
Obviously, the relation (\ref{v2}) implies an isomorphism from the
Hilbert space of the stratification onto the closed linear span of
the orthogonal polynomials with respect to the measure $\mu$.

From orthonormality of the unit vectors $\ket{\phi_i}$ given in
Eq.(\ref{unitv}) (with $\ket{\phi_0}$ as unit vector assigned to
the reference node) we have
\begin{equation}\label{ortpo}
\delta_{ij}=\langle\phi_i|\phi_j\rangle=\int_{R}P_i(x)P_j(x)\mu(dx),
\end{equation}
where, we have used the equation (\ref{phi}).
 Now, by substituting
(\ref{phi}) in (\ref{QD2}) and rescaling $P_k$ as
$Q_k=\sqrt{\omega_1\ldots\omega_k}P_k$, the spectral distribution
$\mu$ under question will be characterized by the property of
orthonormal polynomials $\{Q_k\}$ defined recurrently by
$$ Q_0(x)=1, \;\;\;\;\;\
Q_1(x)=x,$$
\begin{equation}\label{op}
xQ_k(x)=Q_{k+1}(x)+\alpha_{k}Q_k(x)+\omega_kQ_{k-1}(x),\;\;\ k\geq
1.
\end{equation}
If such a spectral distribution is unique, the spectral
distribution $\mu$ is determined by the identity
\begin{equation}\label{sti}
G_{\mu}(x)=\int_{R}\frac{\mu(dy)}{x-y}=\frac{1}{x-\alpha_0-\frac{\omega_1}{x-\alpha_1-\frac{\omega_2}
{x-\alpha_2-\frac{\omega_3}{x-\alpha_3-\cdots}}}}=\frac{Q_{d}^{(1)}(x)}{Q_{d+1}(x)}=\sum_{l=0}^{d}
\frac{\gamma_l}{x-x_l},
\end{equation}
where, $x_l$ are the roots of the polynomial $Q_{d+1}(x)$.
$G_{\mu}(x)$ is called the Stieltjes/Hilbert transform of spectral
distribution $\mu$ and polynomials $\{Q_{k}^{(1)}\}$ are defined
recurrently as
$$Q_{0}^{(1)}(x)=1, \;\;\;\;\;\
    Q_{1}^{(1)}(x)=x-\alpha_1,$$
\begin{equation}\label{oq}
xQ_{k}^{(1)}(x)=Q_{k+1}^{(1)}(x)+\alpha_{k+1}Q_{k}^{(1)}(x)+\omega_{k+1}Q_{k-1}^{(1)}(x),\;\;\
k\geq 1,
\end{equation}
respectively. The coefficients $\gamma_l$ appearing in (\ref{sti})
are calculated as
\begin{equation}\label{Gauss}
\gamma_l:=\lim_{x\rightarrow x_l}[(x-x_l)G_{\mu}(x)]
\end{equation}
Now let $G_{\mu}(z)$ is known, then the spectral distribution
$\mu$ can be determined in terms of $x_l, l=1,2,...$ and Gauss
quadrature constants $\gamma_l, l=1,2,... $ as
\begin{equation}\label{m}
\mu=\sum_{l=0}^d \gamma_l\delta(x-x_l)
\end{equation}
(for more details see Refs. \cite{obh,st,tsc,obah}).
\section{PST of a qudit over antipodes of pseudo distance-regular networks}
\subsection{State Transference in $d$-dimensional
Quantum Systems} A $d$-dimensional quantum system associated with
a simple, connected, finite graph $G=(V,E)$ is defined by
attaching a $d$-level particle to each vertex of the graph so that
with each vertex $i\in V$ one can associate a Hilbert space
${\mathcal{H}}_i\simeq {\mathcal{C}}^d$. The Hilbert space
associated with $G$ is then given by
\begin{equation}
{\mathcal{H}}_G = \otimes_{_{i\in V}}{\mathcal{H}}_i =
({\mathcal{C}}^d)^{\otimes N},
\end{equation}
where $N:=|V|$ denotes the total number of vertices (sites) in
$G$.

Then, the quantum state transfer protocol involves two steps:
initialization and evolution. First, a quantum state
$$\ket{\psi}_A=a_0\ket{0}_A+\sum_{\nu=1}^{d-1}a_{\nu}\ket{\nu}_A\in
{\mathcal{H}}_A$$ (with $a_{\nu}\in \mathcal{C}$ and
$\sum_{\nu=0}^{d-1}|a_{\nu}|^2=1$) to be transmitted is created.
The state of the entire spin system after this step is given by
\be\label{eq1}\ket{\psi(t=0)}=\ket{\psi_A}\otimes\ket{0...00_B}=a_0\ket{0_A}\otimes\ket{0...00_B}+a_1\ket{1_A}\otimes\ket{0...00_B}+...+a_{d-1}\ket{(d-1)_A}\otimes\ket{0...00_B}.\ee
Then, the network couplings are switched on and the whole system
is allowed to evolve under $U(t)=e^{-iHt}$ for a fixed time
interval, say $t_0$. From the fact that
$H\ket{0_A}\otimes\ket{0...00_B}=0$, the final state at time $t_0$
will be\be\label{st} \ket{\psi(t_0)} =
a_0\ket{0_A0...00_B}+\sum_{\nu=1}^{d-1}a_{\nu}\{\sum_{k=1}^Nf^{(\nu)}_{kA}(t_0)\ket{0\ldots
\underbrace{\nu}_{k-th}0...0}\}, \ee where
$f^{(\nu)}_{kA}(t_0):=\langle
0...0\underbrace{\nu}_{k-th}0...0|e^{-iHt_0}|\nu_A0...0\rangle$
for $k=1,2,...,N$; $\nu=1,\ldots,d-1$. In order to transfer the
state $\ket{\psi_A}$ to the site $B$ perfectly (in order to PST is
achieved), the following conditions must be fulfilled
\be\label{eq3} |f^{(\nu)}_{AB}(t_0)|=1\;\;\ \mbox{for}\;\
\nu=1,2,...,d-1 \;\ \mbox{and}\;\  \mbox{some}\;\ 0<t_0<\infty\ee
which can be interpreted as the signature of perfect communication
(or PST) between $A$ and $B$ in time $t_0$. The effect of the
modulus in (\ref{eq3}) is that the state (\ref{st}) will be
$$\ket{\psi(t_0)}=a_0\ket{0_A0...0_B}+\sum_{\nu=1}^{d-1}e^{i\phi_{\nu}}a_{\nu}\ket{0_A0...0}\otimes \ket{\nu}_B$$
so, the state at $B$, after transmission, will no longer be
$\ket{\psi}_A$, but will be of the form \be
a_0\ket{0}+\sum_{\nu=1}^{d-1}e^{i\phi_{\nu}}a_{\nu}\ket{\nu}_B.
\ee The phase factors $e^{i\phi_{\nu}}$ for $\nu=1,2,...,d-1$ are
independent of $a_0,\ldots,a_{d-1}$ and will thus be known
quantities for the graph, which one can correct for with
appropriate phase gates.

The model we will consider is a pseudo distance-regular network
consisting of $N$ sites labeled by $\{1,2, ... ,N\}$ and diameter
$D$. In Ref. \cite{PST}, we introduced the PST of a qubit in terms
of the $SU(2)$ generators. Let us now consider a state with $d$
levels. First, we prepare the generators for $SU(d)$ systems and
thereby introduce the Hamiltonians for a qudit system. The
generators of $SU(d)$ group may be conveniently constructed by the
elementary matrices of $d$ dimension, $\{e_{pq}| p,q\in \{1, . . .
,d\}\}$. The elementary matrices are given by \be\label{ei}
(e_{pq})_{ij}=\delta_{ip}\delta_{jq},\;\ 1\leq i,j\leq d ; \;\;\
e_p:=e_{pp}.\ee which are matrices with one matrix element equal
to unity and all others equal to zero. These matrices satisfy the
commutation relation
$$[e_{pq},e_{rs}]=\delta_{sp}e_{rq}-\delta_{qr}e_{ps}.$$ There are
$d(d-1)$ traceless matrices
$$\hspace{-3.2cm}\lambda^+_{pq}=e_{pq}+e_{qp},$$ \be\label{generators}
\lambda^-_{pq}=\frac{1}{i}(e_{pq}-e_{qp}) ;   \;\;\ 1\leq p<q\leq
d
 \ee
 which are the
off-diagonal generators of the $SU(d)$ group. The $d-1$ additional
traceless matrices \be\label{generators1}
H_{m}=\frac{2}{\sqrt{2m(m+1)}}\{\sum_{k=1}^me_{k}-me_{m+1}\} ;
\;\;\ m=1,2,...,d-1
 \ee
are the diagonal generators so that we obtain a total of $d^2-1$
generators. $SU(2)$ generators are, for instance, given as
$\sigma_x=\lambda^+_{12}=e_{21}+e_{12},
\sigma_y=\lambda^-_{21}=-i(e_{12}-e_{21})$ and
$\sigma_z=H_1=e_1-e_2$.

We now assume that at time $t=0$, the qudit in the first (input)
site of the network is prepared in the state $\ket{\psi_{in}}$. We
wish to transfer the state to the $N$th (output) site of the
network with unit efficiency after a well-defined period of time.
As regards in the above argument, we choose the standard basis
$e_i=\ket{i}$, $i=0,1,...,d-1$ for an individual qudit, and assume
that initially all particles are in the state $\ket{0}$; i.e., the
network is in the state $|\b{0}\rangle=|0_A00...00_B\rangle$.
Then, we consider the dynamics of the system to be governed by the
quantum-mechanical Hamiltonian
\begin{equation}\label{H}
H_G =\frac{1}{2} \sum_{m=0}^DJ_{m}P_m(\frac{1}{2}\sum_{i\sim
j}{\mathbf{\vec{\lambda}}}_i\cdot
{\mathbf{\vec{\lambda}}}_j+[\kappa_{max}-|E|(\frac{d-1}{d})]I_{d^N}),
\end{equation}
where, ${\mathbf{\vec{\lambda}}}_i$ is a $d^2-1$ dimensional
vector with generators of $SU(d)$ as its components acting on the
one-site Hilbert space ${\mathcal{H}}_i$, $J_{m}$ is the coupling
strength between the reference site $1$ and all of the sites
belonging to the $m$-th stratum with respect to $1$, and $P_m$'s
are polynomials given in (\ref{phi}) which are obtained using
three term recursion relations (\ref{op}) and the fact that
$P_m=\frac{1}{\sqrt{\omega_1\omega_2\ldots\omega_m}}Q_m$. As it is
seen from the Eq. (\ref{H}), the terms of the hamiltonian for
$m\geq 1$ are nonlinear functions of $\sum_{i\sim
j}{\mathbf{\vec{\lambda}}}_i\cdot {\mathbf{\vec{\lambda}}}_j$.

In the following we note that the term
$H_{ij}:={\mathbf{\vec{\lambda}}}_i\cdot
{\mathbf{\vec{\lambda}}}_j$ in the hamiltonian (\ref{H}),
restricted to the one particle subspace (the subspace of the full
Hilbert space spanned by the states with only one site excited),
is related to the adjacency matrix of the corresponding graph. To
do so, we write $H_{ij}$ as follows
\begin{equation}\label{hij'}
H_{ij} = \sum_{1\leq p<q\leq d}(\lambda^{+(i)}_{pq}\otimes
\lambda^{+(j)}_{pq}+\lambda^{-(i)}_{pq}\otimes
\lambda^{-(j)}_{pq})+\sum_{m=1}^{d-1}H^{(i)}_m\otimes H^{(j)}_m.
\end{equation}

Before we proceed, one should notice that we have \be\label{eq}
e_{pq}\otimes e_{rs}=e_{(p-1)d+r,(q-1)d+s} \ee Then, from the fact
that $$\lambda^{+(i)}_{pq}\otimes
\lambda^{+(j)}_{pq}+\lambda^{-(i)}_{pq}\otimes
\lambda^{-(j)}_{pq}=2(e^{(i)}_{pq}\otimes
e^{(j)}_{qp}+e^{(i)}_{qp}\otimes e^{(j)}_{pq}),$$ and using the
notation
$$(m,n)\equiv m+(n-1)d,$$ one can obtain \be\label{eq'}\sum_{1\leq
p<q\leq d}(\lambda^{+(i)}_{pq}\otimes
\lambda^{+(j)}_{pq}+\lambda^{-(i)}_{pq}\otimes
\lambda^{-(j)}_{pq})=2\sum_{1\leq p<q\leq
d}[e_{(q,p),(p,q)}+e_{(p,q),(q,p)}]\ee

Now, we evaluate the term $\sum_{m=1}^{d-1}H_m\otimes H_m$ in
(\ref{hij'}) in terms of the elementary matrices $e_{pq}$ as
follows: First we note that \be\label{eq''}H_m\otimes
H_m=\frac{2}{m(m+1)}\{\sum_{p=1}^m\sum_{p'=1}^me_{(p',p)}-m\sum_{p=1}^m[e_{(m+1,p)}+e_{(p,m+1)}]+m^2e_{(m+1,m+1)}\}.\ee
The Eq. (\ref{eq''}) can be rewritten as follows $$H_m\otimes
H_m=\{\frac{2}{m(m+1)}\sum_{p=1}^me_{(p,p)}+\frac{2m}{m+1}e_{(m+1,m+1)}\}+\{\frac{2}{m(m+1)}\sum_{p,p'=1;p\neq
p'}^m e_{(p',p)}-$$
\be\label{eq'''}\frac{2}{m+1}\sum_{p=1}^m[e_{(m+1,p)}+e_{(p,m+1)}]\}.\ee
Then, one can show that \be\label{eq4}\sum_{m=1}^{d-1}H_m\otimes
H_m=2\sum_{p=1}^de_{(p,p)}-\frac{2}{d}I,\ee for proof see Appendix
$A$.

Therefore, by using (\ref{eq'}) and (\ref{eq4}), $H_ {ij}$ in
(\ref{hij'}) is written as follows \be\label{finalH}
H_{ij}=2\sum_{1\leq p<q\leq
d}[e_{(q,p),(p,q)}+e_{(p,q),(q,p)}]+2\sum_{p=1}^de_{(p,p)}-\frac{2}{d}I.\ee

Now, one should notice that the permutation matrix which permutes
two qudits, in terms of the elementary basis $e_{pq}$ can be
written as \be
P=\sum_{p,q=1}^de_{(p,q),(q,p)}=\sum_{p=1}^de_{(p,p)}+\sum_{1\leq
p<q\leq d}^d[e_{(p,q),(q,p)}+e_{(q,p),(p,q)}].\ee Then, the
Eq.(\ref{finalH}) takes the following form \be\label{finalH'}
H_{ij}=2P_{ij}\otimes I_{d^{N-2}}-\frac{2}{d}I_{d^N},\ee where,
$P_{ij}$ denotes the permutation operator which permutes $i$-th
and $j$-th sites and $I_{d^N}$ is $d^N\times d^N$ identity matrix,
where $N$ is the number of vertices or sites ($N:=|V|$).

Now, we denote a state in which the $i$-th site has been exited to
the level $\nu$ by
$\ket{\nu_i}\equiv\ket{0\ldots\ldots0\underbrace{\nu}_i0\ldots0}$.
Then, the permutation operators $P_{ij}$ only changes this state
so that the number and type of excited local states is fixed. This
implies that the hamiltonian $H_G$ can be diagonalized in each
subspace $S^{(\nu)}$ spanned by the vectors $\ket{\nu_i}$,
$i=1,\ldots, N$, for $\nu=1,...,d-1$.

We will refer to the states with only one site excited as one
particle states and the subspace spanned by these vectors comprise
the one-particle sector of the full Hilbert space. Then, the whole
one particle subspace $S$ can be written as
$$S=S^{(1)}\oplus S^{(2)}\oplus\ldots\oplus S^{(d-1)}.$$
In the other words, in $d^N$ dimensional Hilbert space, we deal
with $d-1$ one particle subspaces (recall that, each of these
subspaces has dimension $N$). In the case of PST of a qubit, we
have only one one-particle subspace of dimension $N$.

One can easily show that, the operator $\sum_{_{i\sim j}}P_{ij}$
restricted to the one particle subspace $S^{(l)}$, can be related
to the adjacency matrix $A$, as follows
\begin{equation}\label{P}
\sum_{_{i\sim
j}}P_{ij}-\sum_{i=1}^N[\kappa_{\mbox{max}}-\kappa(i)]H^{(i)}_{l}=A+(|E|-\kappa_{\mbox{max}})I_N,
\end{equation}
where $\kappa_{\mbox{max}}:=\mbox{max}\{\kappa(i),\;\
i=1,2,\ldots, N\}$, $|E|$ is the number of the edges of the graph,
and $H^{(i)}_{l}$ is the projection operator $I\otimes \ldots
\otimes I\otimes \underbrace{H_l}_i\otimes I\ldots I$ with $H_l$
defined as in (\ref{generators1}). For regular graphs, where we
have $\kappa(i)\equiv \kappa$ for all $i=1,2,\ldots, N$, the
equation (\ref{P}) reads as
\begin{equation}\label{Preg}
\sum_{_{i\sim j}}P_{ij}=A+\kappa(\frac{N-2}{2}) I_N,
\end{equation}
in which we have substituted $|E|=\frac{N\kappa}{2}$.

Then, by using (\ref{finalH'}) and (\ref{P}), the hamiltonian in
(\ref{H}) restricted to each one particle subspace $S^{(\nu)}$,
for $\nu=1,2,\ldots,d-1$, can be written in terms of the adjacency
matrix $A$, as follows
\begin{equation}\label{HA}
H_G=\frac{1}{2}\sum_{m=0}^DJ_mP_m(\sum_{_{i\sim
j}}P_{ij}-\frac{|E|}{d}I_N+[\kappa_{max}-|E|(\frac{d-1}{d})]I_N)=\sum_{m=0}^DJ_mP_m(A).
\end{equation}

For the purpose of the perfect transference of a qudit, we
consider pseudo-distance-regular graphs with
$\kappa_D=|\Gamma_D(o)|=1$, i.e., the last stratum of the graph
contains only one site. Then, we impose the constraints that the
amplitudes
$\langle\phi^{(\nu)}_i|e^{-iHt_0}|\phi^{(\nu)}_0\rangle$ be zero
for all $i=0,1,...,D-1$ and $\nu=1,2,\ldots,d-1$, and
$\langle\phi^{(\nu)}_D|e^{-iHt_0}|\phi^{(\nu)}_0\rangle=e^{i\theta}$,
where $\theta$ is an arbitrary phase. Recall that we have
$$|\phi^{(\nu)}_0\rangle=\ket{\nu0\ldots0},\;\ |\phi^{(\nu)}_i\rangle=P_i(A)|\phi^{(\nu)}_0\rangle;\;\;\ \nu=1,2,\ldots,d-1.$$

Therefore, the amplitudes
$\langle\phi^{(\nu)}_i|e^{-iHt_0}|\phi^{(\nu)}_0\rangle$, for
$i=0,1,\ldots,D$ must be evaluated. One can easily show that these
amplitudes are independent of the value of $\nu$, i.e., it
suffices to evaluate them for one choice of $\nu$, say $\nu=1$.

As it has been shown in Ref. \cite{PST}, the above mentioned
constraints (by choosing $\nu=1$, the other choices give the same
results),
are equivalent to the following ones:\\
$$\sum_{k=0}^D\gamma_kP_i(x_k)e^{-2it_0\sum_{m=0}^DJ_mP_m(x_k)}=0, \;\;\ i=0,1,...,D-1.$$
Denoting $e^{-2it_0\sum_{m=0}^DJ_mP_m(x_k)}$ by $\eta_k$, the
above constraints are rewritten as follows
$$
\sum_{k=0}^DP_i(x_k)\eta_k\gamma_k=0,\;\;\ i=0,1,...,D-1,$$
\begin{equation}\label{Cons.}
\sum_{k=0}^DP_D(x_k)\eta_k\gamma_k=e^{i\theta}.
\end{equation}
The Eq.(\ref{Cons.}) can be written as
\begin{equation}\label{Cons.1}
{\mathrm{P}}\left(\begin{array}{c}
  \eta_0\gamma_0 \\
  \eta_1\gamma_1 \\
  \vdots \\
  \eta_D\gamma_D
\end{array}\right)=\left(\begin{array}{c}
  0 \\
  \vdots\\
  0 \\
  e^{i\theta}
\end{array}\right),
\end{equation}
with ${\mathrm{P}}_{ij}=P_i(x_j)$. From the orthogonality relation
(\ref{ortpo}) and using (\ref{m}) , one can obtain
$$\delta_{ij}=\langle\phi_i|\phi_j\rangle=\sum_{l=0}^{D}P_i(x_l)\gamma_lP_j(x_l)\;\
\rightarrow\;\ {\mathrm{P}}W{\mathrm{P}}^t=I\;\ \rightarrow\;\
{\mathrm{P}}W^{1/2}({\mathrm{P}}W^{1/2})^t=I\;\ \rightarrow $$
$$({\mathrm{P}}W^{1/2})^{-1}=({\mathrm{P}}W^{1/2})^t\;\ \rightarrow\;\ {\mathrm{P}}^{-1}=W{\mathrm{P}}^t,$$
with $W:=diag(\gamma_0,\gamma_1,\ldots,\gamma_{D})$. Therefore,
${\mathrm{P}}$ is invertible and the Eq. (\ref{Cons.1}) can be
rewritten as
\begin{equation}\label{Cons.1'}
\left(\begin{array}{c}
  \eta_0\gamma_0 \\
  \eta_1\gamma_1 \\
  \vdots \\
  \eta_D\gamma_D
\end{array}\right)={\mathrm{P}}^{-1}\left(\begin{array}{c}
  0 \\
  \vdots\\
  0 \\
  e^{i\theta}
\end{array}\right)=W{\mathrm{P}}^{t}\left(\begin{array}{c}
  0 \\
  \vdots\\
  0 \\
  e^{i\theta}
\end{array}\right).
\end{equation}

The above equation implies that $\eta_k\gamma_k$ for $k=0,1,...,D$
are the same as the entries in the last column of the matrix
${\mathrm{P}}^{-1}=WP^t$ multiplied with the phase $e^{i\theta}$,
i.e., for the purpose of PST, the following equations must be
satisfied \be\label{result}
\eta_k\gamma_k=\gamma_ke^{-2it_0\sum_{m=0}^DJ_mP_m(x_k)}=e^{i\theta}{(W{\mathrm{P}}^{t})}_{kD}\;\
, \;\;\ \mbox{for} \;\ k=0,1,...,D .\ee One should notice that,
the Eq.(\ref{result}) can be rewritten as  \be\label{res'}
(J_0,J_1,\ldots,J_D)=-\frac{1}{2t_0}[\theta+(2l_0+f(0))\pi,\theta+(2l_1+f(1))\pi,\ldots,\theta+(2l_D+f(D))\pi](W{\mathrm{P}}^{t}),\ee
or \be\label{res''}
J_k=-\frac{1}{2t_0}\sum_{m=0}^D[\theta+(2l_m+f(m))\pi](W{\mathrm{P}}^{t})_{mk},\ee
where $l_k$ for $k=0,1,\ldots, D$ are integers and $f(k)$ is equal
to $0$ or $1$ (we have used the fact that $\gamma_k$ and
${(W{\mathrm{P}}^{t})}_{kD}$ are real for $k=0,1,\ldots, D$, and
so we have $\gamma_k=|{(W{\mathrm{P}}^{t})}_{kD}|$). The result
(\ref{res''}) gives an explicit formula for suitable coupling
constants so that PST between the first node ($\ket{\phi_0}$) and
the opposite one ($\ket{\phi_D}$) can be achieved.
\section{Examples of pseudo-distance-regular networks}
\textbf{1. The networks $G_n$}\\
The networks $G_n$ presented in \cite{fgg}, consist of two
balanced binary trees of height $n$ with the $2^n$ leaves of the
left tree identified with the $2^n$ leaves of the right tree in
the simple way shown in Fig. $1$ (for $n = 2$). The number of
vertices in $G_n$ is $2^{n+1}+ 2^n - 2$. For the purpose of PST
over $G_n$, we prepare the initial state to be transferred, at the
left root of the graph and wants to calculate the suitable
strength coupling constants so that the probability of the
presence of the initial state at the right root be equal to one
for some finite time $t_0$. One can show that $G_n$ is a
pseudo-distance-regular graph with $(2n + 1)$ strata, where
stratum $j$ consists of $2^j$ vertices for $j = 1, 2, ..., n + 1$
and $2^{(2n+1-j)}$ for $j = n + 1, ..., 2n + 1$. Therefore, its QD
parameters are
$$\alpha_i =0, \;\ i=0,1,\ldots, 2n; \;\ \omega_i= 2,\;\ i=1,2,\ldots ,2n.$$
Then one can show that \cite{js,js1} the polynomials $P_i(x)$ are
given by \be
P_i(x)=\frac{1}{\sqrt{2^i}}Q_i(x)=U_i(\frac{x}{2\sqrt{2}}),\ee
where $U_i$'s are the Tchebishef polynomial of the second kind.
Then, $x_l$'s (the roots of the Tchebishef polynomial
$Q_{2n+1}(x)=\sqrt{2^{2n+1}}U_{2n+1}(\frac{x}{2\sqrt{2}})$) and
the coefficients $\gamma_l$ are given by $$\hspace{-3cm}
x_l=2\sqrt{2}\cos \frac{(l+1)\pi}{2(n+1)},$$ \be
\gamma_l=\frac{(-1)^{l+1}}{n+1}\sin\frac{(l+1)\pi}{2(n+1)}\sin\frac{(2n+1)(l+1)\pi}{2(n+1)},\;\
l=0,1,\ldots,2n.\ee Therefore, we have
${\mathrm{P}}_{ij}=U_i(\frac{x_j}{2\sqrt{2}})=U_i(\cos\frac{\pi
(j+1)}{2(n+1)})$ or equivalently \be\label{push}
\mathrm{P}=\left(\begin{array}{cccc}
  1 & 1 & \ldots & 1 \\
  U_1(\frac{\pi}{2(n+1)}) & U_1(\cos\frac{\pi
}{n+1}) & \ldots & U_1(\cos\frac{(2n+1)\pi
}{2(n+1)}) \\
  \vdots & \vdots & \ldots & \vdots \\
  U_{2n}(\frac{\pi}{2(n+1)}) & U_{2n}(\cos\frac{\pi
}{n+1}) & \ldots & U_{2n}(\cos\frac{(2n+1)\pi
}{2(n+1)}) \\
\end{array}\right).
\ee Furthermore, as it is seen from (\ref{push}), the matrix
$\mathrm{P}$ is a polynomial transformation \cite{Puschel}. By
using the result (\ref{res''}), we obtain \be\label{resgn}
J_k=-\frac{1}{2t_0}\sum_{m=0}^{2n}
\frac{(-1)^{m+1}}{n+1}\sin\frac{(m+1)\pi}{2(n+1)}\sin\frac{(2n+1)(m+1)\pi}{2(n+1)}[\theta+(2l_m+f(m))\pi]U_k(\cos
\frac{(m+1)\pi}{2(n+1)}),\ee for $k=0,1, \ldots,2n$.

In the following we consider the case $n=2$ in detail. In this
case, we have
$$x_0=\sqrt{6},\;\ x_1=\sqrt{2},\;\ x_2=0,\;\ x_3=-\sqrt{2},\;\ x_4=-\sqrt{6}; $$
$$\gamma_0=\gamma_4=\frac{1}{12},\;\ \gamma_1=\gamma_3=\frac{1}{4},\;\ \gamma_2=\frac{1}{3}.$$
Then by using the equations (\ref{push}) and (\ref{Cons.1'}), we
have
$$\mathrm{P}=\left(\begin{array}{ccccc}
  1 & 1 & 1 & 1 & 1 \\
  \sqrt{3} & 1 & 0 & -1 & -\sqrt{3} \\
  2 & 0 & -1 & 0 & 2 \\
  \sqrt{3} & -1 & 0 & 1 & -\sqrt{3} \\
  1 & -1 & 1 & -1 & 1\\
\end{array}\right)\;\  \rightarrow \;\ {\mathrm{P}}^{-1}=W{\mathrm{P}}^t=\frac{1}{12}\left(\begin{array}{ccccc}
  1 & \sqrt{3} & 2 & \sqrt{3} & 1 \\
  3 & 3 & 0 & -3 & -3 \\
  4 & 0 & -4 & 0 & 4 \\
  3 & -3 & 0 & 3 & -3 \\
  1 & -\sqrt{3} & 2 & -\sqrt{3} & 1\\
\end{array}\right).$$
Now, using the Eq. (\ref{resgn}), one can obtain the following
suitable coupling constants in order to PST between the most left
and the most right nodes be achieved:
$$J_0=-\frac{\theta+2\pi/3}{2t_0},\;\ J_1=J_3=\frac{\pi}{4\sqrt{3}t_0},\;\ J_2=-J_4=-\frac{\pi}{6t_0}.$$
\textbf{2. The modified $G_n$-type networks}\\
This type of networks consist of two balanced binary trees of
height $n+1$. For the graphs $G_n$, set entrance and exit to the
left root and the right root of the trees, respectively (see Fig.
$2$ for $n=2$). These networks are pseudo distance regular
networks which are characterized by the following intersection
array
$$\{b_0,b_1,\ldots, b_{2n-1};c_1,c_2,\ldots,c_{2n}\}=\{3,\underbrace{2,2,\ldots,2}_{n-1},\underbrace{1,1,\ldots,1}_{n};\underbrace{1,1,\ldots,1}_{n},\underbrace{2,2,\ldots,2}_{n-1},3\}.$$
Then by using (\ref{intersec'}) and (\ref{omegal0}), one can
obtain
$$\kappa\equiv \kappa_1=3,\;\ \kappa_i=\kappa_{2n-i}=3\times 2^{i-1},\;\ i=1,2,\ldots,n\;\ \kappa_{2n}=1,$$
$$\alpha_i=0,\;\ i\neq n+1,\;\  \alpha_{n+1}=1 ;\;\ \omega_1=\omega_{2n}=3,\;\ \omega_i=2,\;\ i=2,\ldots,2n-1.$$
Now, by using the QD parameters $\alpha_i$ and $\omega_i$, and the
recursion relations (\ref{op}) and (\ref{oq}), one can obtain the
Stieltjes function for any given $n$ and spectral distribution
$\mu(x)$ for the modified $G_n$-type network. In the following we
consider the case $n=2$ in detail.\\
In this case, we have
$$\alpha_i=0,\;\ i=0,1,2,4,\;\  \alpha_{3}=1 ;\;\ \omega_1=\omega_{4}=3,\;\ \omega_2=\omega_3=2.$$
Then by using the recursion relations (\ref{op}) and (\ref{oq}),
one can obtain the stieltjes function as
$$G_{\mu}(x)=\frac{x^4-7x^2-x^3+2x+6}{x(x-3)(x^3+2x^2-4x-7)}.$$
By using the recursion relations (\ref{op}), we obtain
\be\label{hadx} \hspace{-1.7cm}Q_1(x)=x,\;\ Q_2(x)=x^2-3,\;\
Q_3(x)=x(x^2-5),\;\ Q_4(x)=x^4-x^3-7x^2+5x+6,\;\
Q_5(x)=x(x^4-x^3-10x^2+5x+21).\ee Then, from the fact that, $x_l$,
$l=0,1,2,3,4$ are roots of $Q_5(x)$ (the denominator of
$G_{\mu}(x)$), we obtain
$$x_0=0,\;\ x_1=3,\;\ x_2\simeq-2.4728,\;\ x_3\simeq-1.4626,\;\ x_4\simeq1.9354;$$
$$\gamma_0=\frac{2}{7},\;\ \gamma_1=\frac{1}{26},\;\
\gamma_2\simeq 0.3101,\;\ \gamma_3\simeq 0.1786,\;\ \gamma_4\simeq
0.1872.$$ By using the Eq.(\ref{op}) and the fact that
${\mathrm{P}_{ij}}=P_{i}(x_j)=\frac{1}{\sqrt{\sqrt{\omega_1\ldots\omega_i}}}Q_i(x_j)$,
we obtain
$$\mathrm{P}\simeq\left(\begin{array}{ccccc}
  1 & 1 & 1 & 1 & 1 \\
  0 & \sqrt{3} & -1.4277 & -0.8445 & 1.1174 \\
  -1.3417 & 0 & 1.3930 & -0.3850 & 0.1492 \\
  0 & 2\sqrt{3} & -0.4594 & 0.6974 & 0.4046 \\
  1 & 2 & 0.5572 & -1.4304 & -0.6270\\
\end{array}\right)\;\  \rightarrow$$
$$
{\mathrm{P}}^{-1}=W{\mathrm{P}}^t\simeq\left(\begin{array}{ccccc}
  0.2869 & -0.0155 & -0.3725 & -0.1982 & 0.2133 \\
  -0.0303 & -0.0157 & 0.0493 & 0.2495 & 0.0965 \\
  0.2509 & -0.1527 & 0.2961 & -0.0807 & 0.1465 \\
  0.0712 & -0.3078 & -0.1565 & 0.2957 & -0.2813 \\
  0.4213 & 0.4916 & 0.1836 & -0.2664 & -0.1750\\
\end{array}\right).$$
Now, using the Eq. (\ref{res''}), one can obtain the following
suitable coupling constants in order to PST between the most left
and the most right nodes be achieved:
$$J_0=-\frac{\theta+0.4925\pi}{2t_0},\;\ J_1=-\frac{0.1838\pi}{2t_0},\;\ J_2=-\frac{0.0271\pi}{2t_0},\;\ J_3=-\frac{0.0293\pi}{2t_0},\;\ J_4=\frac{0.4563\pi}{2t_0}.$$
\textbf{3. Pseudo-distance-regular network derived from Icosahedron network}\\
This network has $12$ nodes with
$$\{b_0,b_1,b_2;c_1,c_2,c_3\}=\{5,2,1;1,2,5\}.$$
Then by using (\ref{intersec'}) and (\ref{omegal0}), one can
obtain
$$\kappa=5,\;\ \kappa_2=5,\;\ \kappa_3=1,$$
$$\alpha_0=0,\;\ \alpha_1=\alpha_2=2,\;\ \alpha_3=0 ;\;\ \omega_1=5,\;\ \omega_2=4,\;\ \omega_3=5.$$
The stratification with respect to the initial node $1$ and the
final node $12$ produces the same strata (see Fig. $3$).
Therefore, the PST between the nodes $1$ and $12$ can be achieved.
To do so, by using the recursion relations (\ref{op}), we obtain
\be\label{had} Q_1(x)=x,\;\ Q_2(x)=x^2-2x-5,\;\
Q_3(x)=x^3-4x^2-5x+10,\;\ Q_4(x)=x^4-4x^3-10x^2+20x+25. \ee Then,
from the fact that, $x_l$, $l=0,1,2,3$ are roots of $Q_4$, we
obtain
$$x_0=-1,\;\ x_1=5,\;\ x_2=\sqrt{5},\;\ x_3=-\sqrt{5}; \;\ \gamma_0=\frac{5}{12},\;\ \gamma_1=\frac{1}{12},\;\ \gamma_2=\gamma_3=\frac{1}{4}.$$
By using the Eq.(\ref{op}) and the fact that
${\mathrm{P}_{ij}}=P_{i}(x_j)=\frac{1}{\sqrt{\sqrt{\omega_1\ldots\omega_i}}}Q_i(x_j)$,
we obtain
$$\hspace{-1.5cm}{\mathrm{P}}=\left(\begin{array}{cccc}
          1 & 1 & 1 & 1 \\
                       -\frac{1}{\sqrt{5}} & \sqrt{5} & 1 & -1 \\
                       -\frac{1}{\sqrt{5}} & \sqrt{5} & -1 & 1 \\
                       1 & 1 & -1 & -1 \\
                     \end{array}
                      \right)\;\ \rightarrow\;\ (\mathrm{P})^{-1}=W{\mathrm{P}}^t=\frac{1}{12}\left(\begin{array}{cccc}
          5 & -\sqrt{5} & -\sqrt{5} & 5 \\
                       1 & \sqrt{5} & \sqrt{5} & 1 \\
                       3 & 3 & -3 & -3 \\
                       3 & -3 & 3 & -3 \\
                     \end{array}
                      \right).$$\\
Now, using the Eq. (\ref{res''}), we obtain the following coupling
constants
$$J_0=-\frac{2\theta+\pi}{4t_0},\;\ J_1=J_2=0,\;\ J_3=\frac{\pi}{4t_0}.$$
\textbf{4. $3$- simplex fractal with decimation number $b=2$}\\
Consider the $(n-1)$-simplex fractal with decimation number $b=2$
\cite{jssps}-\cite{jrest1} such that all of the $n-1$ vertices
$\underbrace{(20...0)}_{n-1}$, $\underbrace{(020...0)}_{n-1}$,
..., $\underbrace{(0...02)}_{n-1}$ are connected to each other
(see Fig. $4$ for $n=4$ ). Then, the number of vertices of
$(n-1)$-simplex fractal is
$N=C^{n-1}_1+C^{n-1}_2=\frac{n(n-1)}{2}$ such that the degree of
each vertex is $\kappa=2(n-2)$. Also, it can be easily shown that
the network has $3$ strata with respect to the reference node
$(200...0)$. For $n=4$, the last stratum contains only one nodes
so that the PST can be considered; In this case, the intersection
array, size of strata and the QD parameters are given by
$$\{b_0,b_1;c_1,c_2\}=\{4,1;1,1,4\},$$
$$\kappa=4,\;\ \kappa_2=1,$$
$$\alpha_0=0,\;\ \alpha_1=2,\;\ \alpha_2=0 ;\;\ \omega_1=\omega_2=4.$$
Then, by using the recursion relations (\ref{op}), we obtain
\be\label{had} Q_1(x)=x,\;\ Q_2(x)=x^2-2x-4,\;\
Q_3(x)=x^3-2x^2-8x. \ee Then, from the fact that, $x_l$,
$l=0,1,2,3$ are roots of $Q_3$, we obtain
$$x_0=0,\;\ x_1=4,\;\ x_2=-2; \;\ \gamma_0=\frac{1}{2},\;\ \gamma_1=\frac{1}{6},\;\ \gamma_2=\frac{1}{3}.$$
By using the Eq.(\ref{op}) and the fact that
${\mathrm{P}_{ij}}=P_{i}(x_j)=\frac{1}{\sqrt{\omega_1\ldots\omega_i}}Q_i(x_j)$,
we obtain
$$\hspace{-1.5cm}{\mathrm{P}}=\left(\begin{array}{ccc}
          1 & 1 & 1  \\
                       0 & 2 & -1 \\
                       -1 & 1 & 1 \\
                       \end{array}
                      \right)\;\ \rightarrow\;\ (\mathrm{P})^{-1}=W{\mathrm{P}}^t=\frac{1}{6}\left(\begin{array}{ccc}
          3 & 0 & -3 \\
                       1 & 2 & 1 \\
                       2 & -2 & 2 \\
                        \end{array}
                      \right).$$
By using the Eq. (\ref{res''}), we obtain the following constants
in order to PST be achieved:
$$J_0=-\frac{2\theta+5\pi/3}{4t_0},\;\ J_1=-\frac{\pi}{3t_0},\;\ J_2=\frac{\pi}{12t_0}.$$
\textbf{5. Pseudo-distance-regular network derived from Hadamard network with $16$ nodes} \\
Consider the pseudo-distance-regular network given in Fig. $5$.
This network is obtained from the Hadamard network with
intersection array $\{4,3,2,1;1,2,3,4\}$. As Fig. $5$ shows, the
network is symmetric with respect to the initial and final
(horizontal) nodes $1$ and $16\in \Gamma_4(1)$ and also with
respect to the initial and final (vertical) nodes $6$ and $11\in
\Gamma_4(6)$. One should notice that stratification of the network
with respect to the nodes $1$ and $16$ produces the same strata.
For stratification with respect to the node $1$ or $16$, we have
$$\kappa=4,\;\ \kappa_2=6,\;\ \kappa_3=4,\;\ \kappa_4=1,$$
$$\alpha_i=0,\;\ i=0,...,4;\;\;\ \omega_1=4,\;\ \omega_2=6,\;\ \omega_3=6,\;\ \omega_4=4.$$
(see Eq. (\ref{QD1})). Then, by using the recursion relations
(\ref{op}), one can obtain \be\label{had'} Q_1(x)=x,\;\
Q_2(x)=x^2-4,\;\ Q_3(x)=x^3-10x,\;\ Q_4(x)=x^4-16x^2+24,\;\
Q_5(x)=x^5-20x^3+64x. \ee Then, from the fact that, $x_l$,
$l=0,1,2,3,4$ are roots of $Q_5$, we obtain
$$\hspace{-1.5cm}x_0=0,\;\ x_1=2,\;\ x_2=-2,\;\ x_3=4,\;\ x_4=-4; \;\;\ \gamma_0=\frac{3}{8},\;\ \gamma_1=\gamma_2=\frac{1}{4},\;\ \gamma_3=\gamma_4=\frac{1}{16}.$$
Then by using the equations (\ref{push}) and (\ref{Cons.1'}), we
have
$$\mathrm{P}=\left(\begin{array}{ccccc}
  1 & 1 & 1 & 1 & 1 \\
  0 & 1 & -1 & 2 & -2 \\
  -\frac{\sqrt{6}}{3} & 0 & 0 & \sqrt{6} & \sqrt{6} \\
  0 & -1 & 1 & 2 & -2 \\
  1 & -1 & -1 & 1 & 1\\
\end{array}\right)\;\  \rightarrow \;\ {\mathrm{P}}^{-1}=W{\mathrm{P}}^t=\frac{1}{16}\left(\begin{array}{ccccc}
  6 & 0 & -2\sqrt{6} & 0 & 6 \\
  4 & 4 & 0 & -4 & -4 \\
  4 & -4 & 0 & 4 & -4 \\
  1 & 2 & \sqrt{6} & 2 & 1 \\
  1 & -2 & \sqrt{6} & -2 & 1\\
\end{array}\right).$$
Now, using the Eq. (\ref{res''}), we obtain the following
constants:
$$J_0=-\frac{\theta+5\pi/8}{2t_0},\;\ J_1=J_3=-\frac{\pi}{8t_0},\;\ J_2=-\frac{3\pi}{8\sqrt{6}t_0},\;\ J_4=\frac{3\pi}{16t_0}.$$
\section{Conclusion}
PST of an arbitrary $d$-level quantum state over antipodes of
pseudo distance-regular networks, was investigated. By using the
spectral analysis techniques and algebraic structures of pseudo
distance-regular graphs an explicit formula for coupling constants
in the Hamiltonians was given so that state of a particular qudit
initially encoded on one site can be evolved freely to the
opposite site without any dynamical control.
\newpage
\setcounter{section}{0} \setcounter{equation}{0}
\renewcommand{\theequation}{A-\roman{equation}}
{\Large{Appendix A}}\\
Proof of the Eq.(3-40):\\
By using the Eq. (\ref{eq'''}), we have
$$\sum_{m=1}^{d-1}H_m\otimes
H_m=\sum_{m=1}^{d-1}\{\frac{2}{m(m+1)}\sum_{p=1}^me_{(p,p)}+\frac{2m}{m+1}e_{(m+1,m+1)}\}+\sum_{m=1}^{d-1}\{\frac{2}{m(m+1)}\sum_{p,p'=1;p\neq
p'}^m e_{(p',p)}-$$
\be\label{a1}\frac{2}{m+1}\sum_{p=1}^m[e_{(m+1,p)}+e_{(p,m+1)}]\}.\ee
We evaluate the first sum in the above equation, the second one
can be evaluated similarly.
$$
\sum_{m=1}^{d-1}\{\frac{2}{m(m+1)}\sum_{p=1}^me_{(p,p)}+\frac{2m}{m+1}e_{(m+1,m+1)}\}=\sum_{m=1}^{d-1}\frac{2}{m(m+1)}\sum_{p=1}^me_{(p,p)}+\sum_{m=1}^{d-1}\frac{2m}{m+1}e_{(m+1,m+1)}
$$
$$=\sum_{p=1}^1e_{(p-1)d+p}+\frac{1}{3}\sum_{p=1}^2e_{(p-1)d+p}+\ldots+\frac{2}{2(d-1)}\sum_{p=1}^{d-1}e_{(p-1)d+p}+e_{d+2}+\frac{4}{3}e_{2d+3}+\ldots+\frac{2(d-1)}{d}e_{d}=$$
$$e_1\sum_{m=1}^{d-1}\frac{2}{m(m+1)}+e_{d+2}\sum_{m=2}^{d-1}\frac{2}{m(m+1)}+\ldots+e_{d-1}\frac{2}{d(d-1)}+e_{d+2}+\frac{4}{3}e_{2d+3}+\ldots+\frac{2(d-1)}{d}e_{d}=$$
$$\sum_{\alpha=1}^{d}[\frac{2(\alpha-1)}{\alpha}+2(\frac{1}{\alpha}-\frac{1}{d})]e_{(\alpha-1)d+\alpha}=2(1-1/d)(e_1+e_{d+2}+\ldots+e_{d^2})=2(1-1/d)\sum_{p=1}^de_{(p-1)d+p},$$
where, we have used the identity
$\sum_{m=\alpha}^{d-1}\frac{2}{m(m+1)}=\sum_{_{m=\alpha}}^{d-1}(\frac{1}{m}-\frac{1}{m+1})=2(\frac{1}{\alpha}-\frac{1}{d})$.

The second sum in (\ref{a1}) can be evaluated as
$$
\sum_{m=1}^{d-1}\{\frac{2}{m(m+1)}\sum_{p,p'=1;p\neq p'}^m
e_{(p',p)}-\frac{2}{m+1}\sum_{p=1}^m[e_{(m+1,p)}+e_{(p,m+1)}]\}=-\frac{2}{d}\sum_{p,p'=1;p\neq
p'}^de_{(p-1)d+p'}.$$ Therefore, we obtain
$$\sum_{m=1}^{d-1}H_m\otimes
H_m=2\sum_{p=1}^{d}e_{(p-1)d+p}-\frac{2}{d}\sum_{p,p'=1}^de_{(p-1)d+p'}=2\sum_{p=1}^{d}e_{(p-1)d+p}-\frac{2}{d}I
. $$

\newpage
{\bf Figure Captions}

{\bf Figure-1:} Shows the network $G_2$.

{\bf Figure-2:} shows the modified $G_2$-type network.

{\bf Figure-3:} Shows the pseudo-distance-regular network derived
from Icosahedron network.

{\bf Figure-4:} Shows the $3$- simplex fractal with decimation
number $b=2$.

{\bf Figure-5:} Shows the pseudo-distance-regular network derived
from Hadamard network with $16$ nodes.
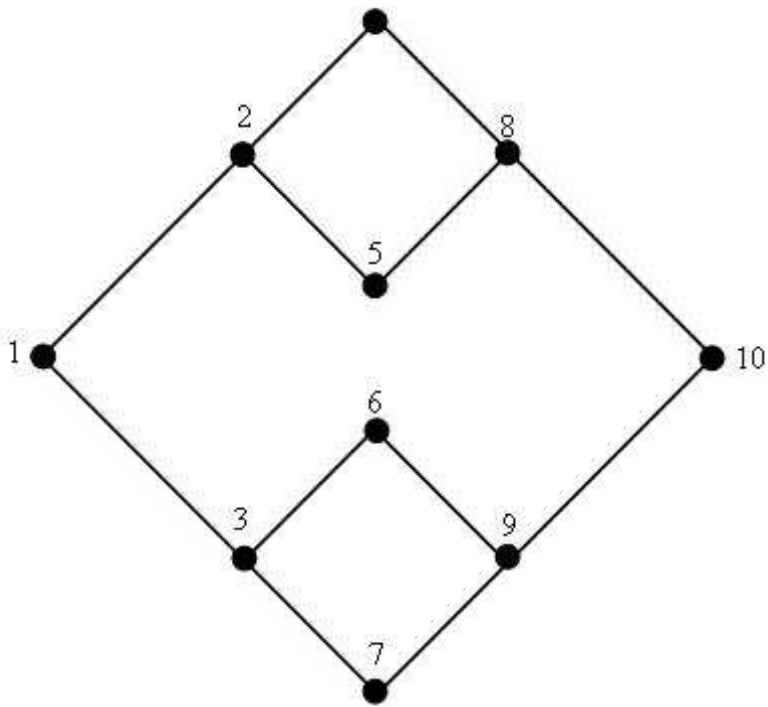
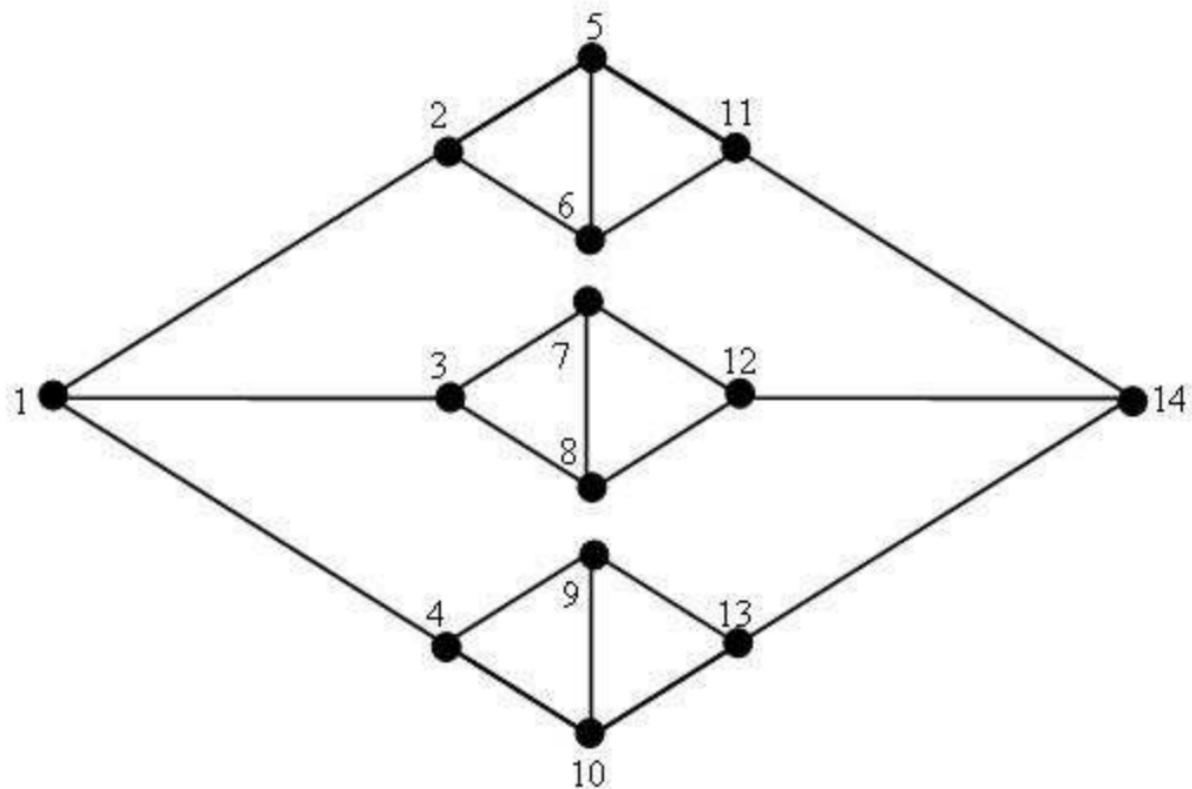
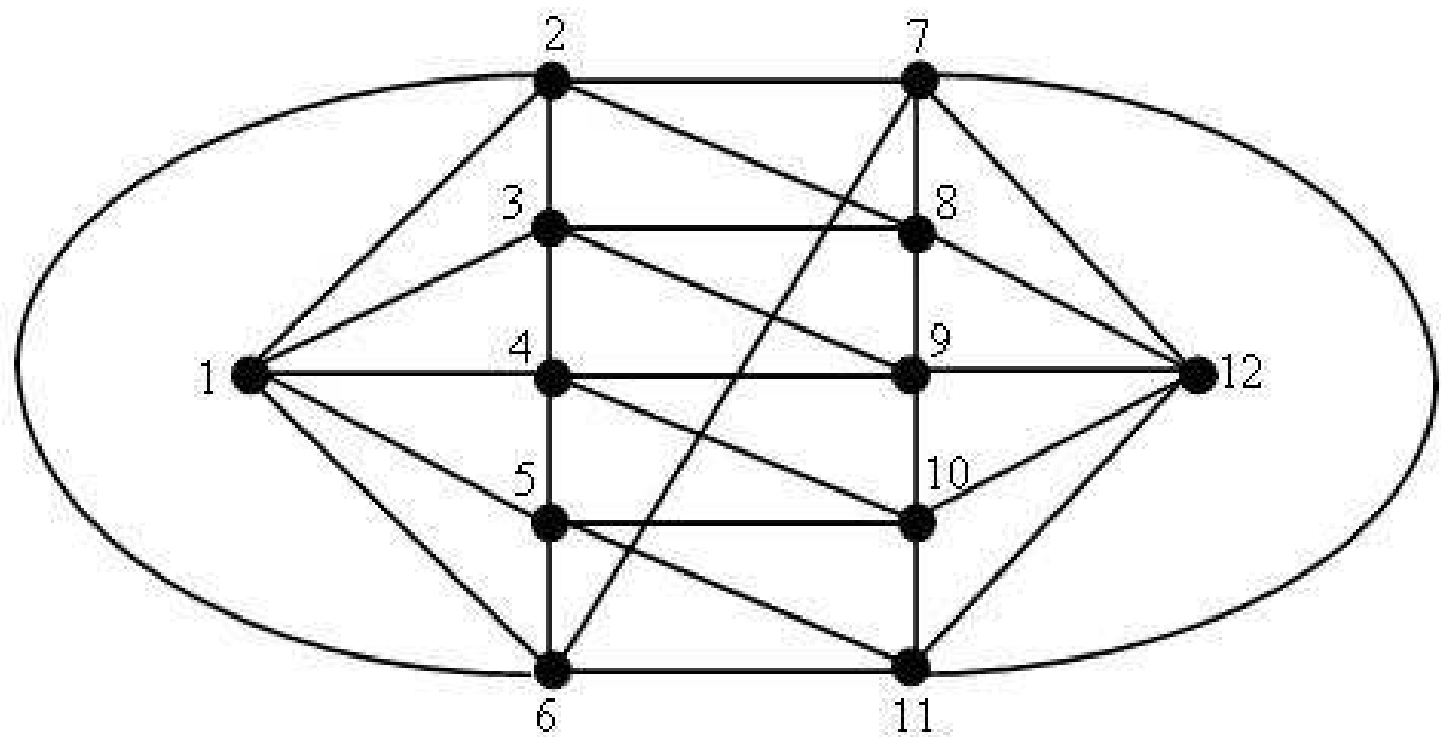
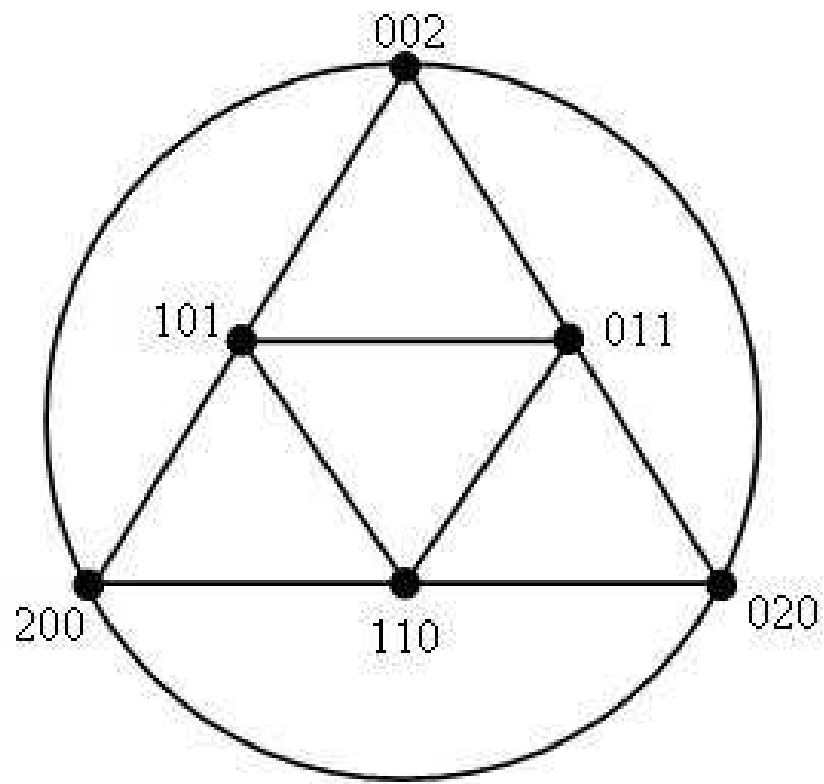
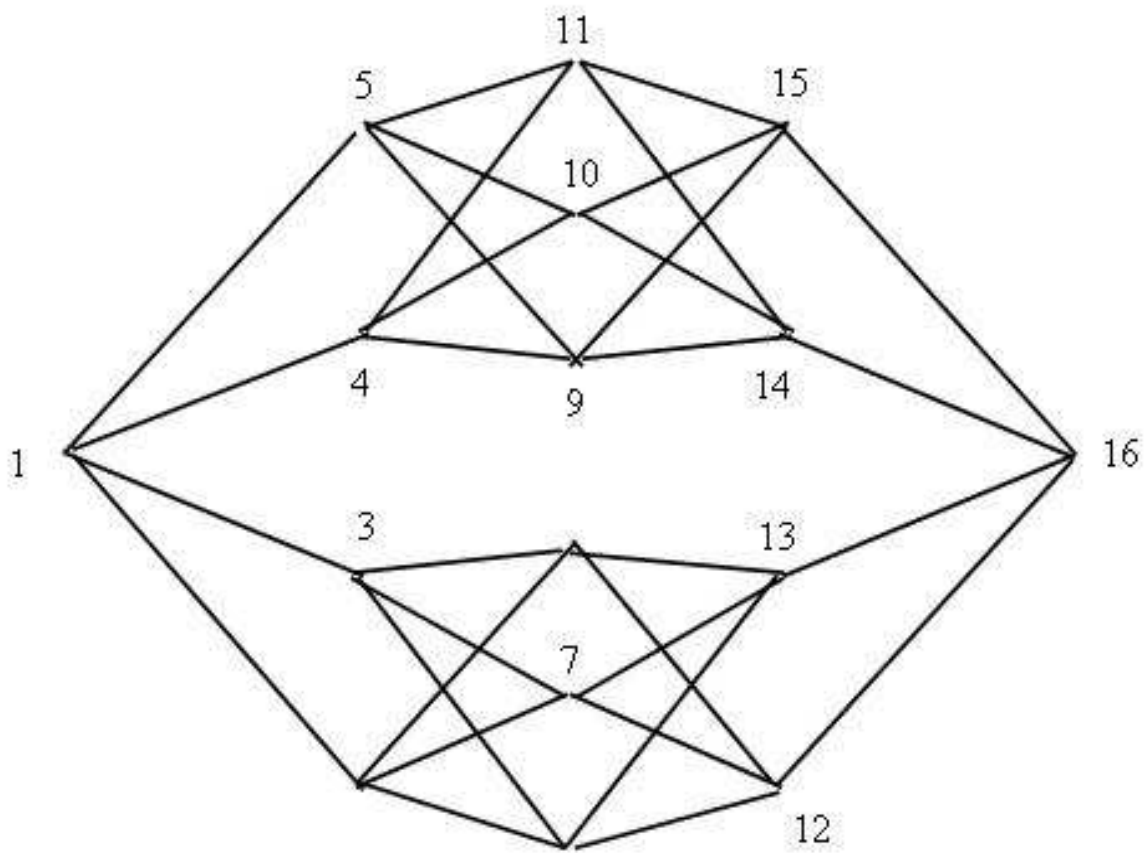
\end{document}